\documentclass[twocolumn]{emulateapj}
\usepackage{lscape}

\shorttitle{The Nascent Red-Sequence at $z \sim 2$}
\shortauthors{Zirm et al.}

\def\spose#1{\hbox to 0pt{#1\hss}}
\def\simlt{\mathrel{\spose{\lower 3pt\hbox{$\mathchar"218$}}
     \raise 2.0pt\hbox{$\mathchar"13C$}}}
\def\simgt{\mathrel{\spose{\lower 3pt\hbox{$\mathchar"218$}}
     \raise 2.0pt\hbox{$\mathchar"13E$}}}

\journalinfo{Accepted for publication in the Astrophysical Journal}

\begin{document}

\title{The Nascent Red-Sequence at ${\mathbf z \sim 2}$\altaffilmark{1}}

\author{Andrew W. Zirm\altaffilmark{2},
  S.A. Stanford\altaffilmark{3,4}, M. Postman\altaffilmark{5},
  R.A. Overzier\altaffilmark{2}, J.P. Blakeslee\altaffilmark{6,7},
  P. Rosati\altaffilmark{8}, J. Kurk\altaffilmark{9},
  L. Pentericci\altaffilmark{10}, B. Venemans\altaffilmark{11},
  G.K. Miley\altaffilmark{12}, H.J.A.~R{\"o}ttgering\altaffilmark{},
  M.~Franx\altaffilmark{}, A.~van der Wel\altaffilmark{2},
  R. Demarco\altaffilmark{2}, W. van Breugel\altaffilmark{3,13}}

\altaffiltext{1}{Based on observations with the NASA/ESA Hubble Space
Telescope, obtained at the Space Telescope Science Institute, which is
operated by the Association of Universities for Research in Astronomy,
Inc., under NASA contract NAS 5-26555}

\altaffiltext{2}{Johns Hopkins University, 3400 N. Charles Street,
Baltimore, MD, 21218; {\tt azirm@pha.jhu.edu}}
\altaffiltext{3}{Institute of Geophysics and Planetary Physics, LLNL, L-413, Livermore, CA 94550}
\altaffiltext{4}{Physics Dept., University California at Davis, One Shields Avenue, Davis, CA 95616}
\altaffiltext{5}{Space Telescope Science Institute, 3700 San Martin Drive, Baltimore, MD 21218}
\altaffiltext{6}{Department of Physics and Astronomy, Washington State University, Pullman, WA 99164}
\altaffiltext{7}{Herzberg Institute of Astrophysics, NRC of Canada, 5071 W. Saanich Road, Victoria, BC, V9E 2E7, Canada}
\altaffiltext{8}{ESO-European Southern Observatory, Karl-Schwarzschild-Strasse 2, D-85748, Garching bei M{\"u}nchen, Germany}
\altaffiltext{9}{Max-Planck-Institut f{\"u}r Astronomie, D-69117 Heidelberg, Germany}
\altaffiltext{10}{Osservatorio Astronomico di Roma, I-00040 Monte Porzio Catone, Italy}
\altaffiltext{11}{IOA, Cambridge University, Madingley Road, Cambridge, CB3 0HA, UK}
\altaffiltext{12}{Leiden Observatory, Leiden University, Postbus 9513,
NL-2300 RA Leiden, The Netherlands}
\altaffiltext{13}{University of California, Merced, PO Box 2039, Merced, CA 95344}

\begin{abstract}
  We present new constraints on the evolution of the early-type galaxy
  color-magnitude relation (CMR) based on deep near-infrared imaging
  of a galaxy protocluster at $z=2.16$ obtained using NICMOS on-board
  the {\it Hubble Space Telescope}.  This field contains a
  spectroscopically confirmed space-overdensity of Lyman-$\alpha$ and
  H-$\alpha$ emitting galaxies which surrounds the powerful radio
  galaxy MRC 1138-262.  Using these NICMOS data we identify a
  significant surface-overdensity ($= 6.2\times$) of red $J_{110} -
  H_{160}$ galaxies in the color-magnitude diagram (when compared with
  deep NICMOS imaging from the HDF-N and UDF).  The optical-NIR colors
  of these prospective red-sequence galaxies indicate the presence of
  on-going dust-obscured star-formation or recently formed ($\simlt
  1.5$ Gyr) stellar populations in a majority of the red galaxies.  We
  measure the slope and intrinsic scatter of the CMR for three
  different red galaxy samples selected by a wide color cut, and using
  photometric redshifts both with and without restrictions on
  rest-frame optical morphology.  In all three cases both the
  rest-frame $U-B$ slope and intrinsic color scatter are considerably
  higher than corresponding values for lower redshift galaxy clusters.
  These results suggest that while some relatively quiescent galaxies
  do exist in this protocluster both the majority of the galaxy
  population and hence the color-magnitude relation are still in the
  process of forming, as expected.
\end{abstract}

\keywords{galaxies: evolution --- galaxies: formation --- galaxies:
  high-redshift --- galaxies: stellar content --- galaxies: clusters:
  individual: MRC 1138-262}

\section{Introduction\label{sec:intro}}

The color-magnitude diagram is a powerful diagnostic of galaxy
evolution and formation.  The presence, as early as $z \sim 1.5$, of a
prominent and low-scatter, `red-sequence' (RS) in galaxy clusters
places useful constraints on the possible evolutionary pathways in
galaxy color and luminosity \citep*{Mullisetal05, Stanfordetal05,
  Stanfordetal06, Belletal04, Faberetal07}.  The red colors of the
primarily early-type RS galaxies are due to the observed filters
spanning the 4000\AA\ spectral break.  The universality and prominence
of the RS in appropriately chosen filters have been used to discover
high-redshift clusters \citep*[e.g.,][]{RCS}.  Moreover, the defining
characteristic of galaxy clusters, i.e., the large numbers of galaxies
all at the same redshift, allows the slope and intrinsic scatter of
the RS to be measured with great precision.  Based on studies of
galaxy clusters at $z < 1.3$, the slope of the RS does not appear to
evolve and therefore is more likely the by-product of the
mass-metallicity relation as observed in local galaxy samples
\citep*[e.g.,][]{Tremontietal04} rather than the result of a mass-age
trend.  The scatter, however, is likely due to the fractional age
differences between the RS galaxies \citep*[e.g.,][]{Blakesleeetal03}.
By constructing a set of model galaxies with different star-formation
histories and timescales it is possible to fit for the mean epoch of
last significant star-formation by matching the intrinsic scatter of
the RS.  Such studies at $z \sim 1$ have derived formation redshifts
of $z_{\rm form} \sim 2.0 - 2.5$ \citep*[e.g.,][]{HalfHubble06,
  vanDokkumvanderMarel07a}.  At redshifts beyond $z \sim 1.5$,
however, the 4000\AA-break moves into the near-infrared and galaxy
clusters, and therefore the RS, have not been observed closer to the
inferred epoch of formation for early-type galaxies.  Hence, to
uncover the younger or forming red-sequence at higher redshifts
requires deep near-infrared imaging of suspected (or, preferably,
confirmed) protocluster fields.

\begin{figure*}[t]
\plotone{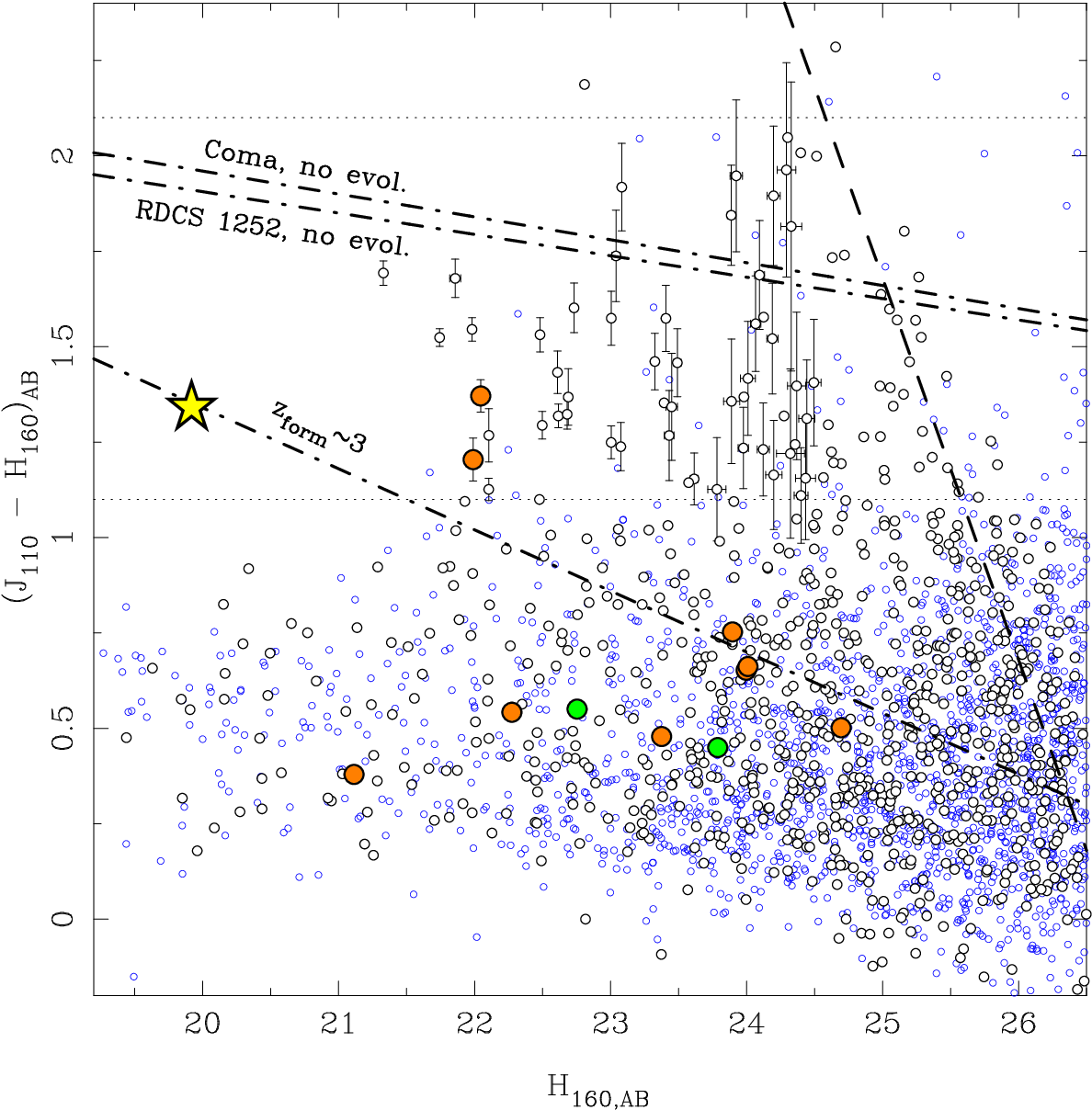}
\caption{$J_{110} - H_{160}$ vs. $H_{160}$ color-magnitude diagram for
  the MRC 1138-262 NICMOS field (open black circles).  The large
  yellow star is the radio galaxy itself.  The blue background points
  are from the NICMOS data of the Ultra Deep Field and Hubble Deep
  Field North.  The deep fields cover 2.5x the area of our
  observations.  Also shown are the spectroscopically confirmed
  $H\alpha$ (orange filled circles) and Lyman-$\alpha$ (green filled
  circles) emitters.  The three dot-dash lines show where the CMRs of
  lower redshift clusters would lie under different assumptions.  The
  top line is the Coma cluster with no evolution, simply redshifted to
  $z=2.16$ and observed through the NICMOS filters.  The next line
  down is the same but for the $z=1.24$ cluster RDCS1252.  Finally, if
  we passively de-evolve RDCS1252 to redshift $z=2.16$ (almost exactly
  two Gyr), assuming a median age for the 1252 galaxies of about 3 Gyr
  (or $z_{\rm form} \sim 3$), we get the third line.\label{fig:CMD}}
\end{figure*}

We have undertaken a NICMOS imaging program to study the red galaxy
population in a protocluster at $z=2.16$.  Broad and narrow-band
imaging, both in the optical and near-infrared, of the field
surrounding the powerful radio galaxy MRC 1138-262 ($z=2.16$) have
identified more than 100 candidate companion galaxies.  This target
served as the proof-of-concept for the successful VLT Large Program
summarized in Venemans et al. (2007)\nocite{Venemansetal07}.  There
are surface-overdensities of both line-emitting candidates
(Lyman-$\alpha$ and H$\alpha$), X-ray point sources, sub-mm selected
galaxies and red optical--near-infrared galaxies
\citep*{Pentericcietal02,KurkPhD,Kurketal04,Kurketal04b,Croftetal05,Stevensetal03}.
Fifteen of the Ly$\alpha$ and 9 of the H$\alpha$ emitters have now
been spectroscopically confirmed to lie at the same redshift as the
radio galaxy.  The $I-K$-selected Extremely Red Objects (EROs; $I-K >
4.3$ Vega) seem concentrated around the RG but have no spectroscopic
redshifts at this time.  However, by obtaining deep images through the
NICMOS $J_{110}$ and $H_{160}$ filters, which effectively span the
4000\AA-break at $z=2.16$, accurate and precise colors and basic
morphological parameters can be measured for the red galaxy
population.  In this paper we present the first results from this
project.  The article is organized as follows: in Section
\S\ref{sec:obs} we describe the data and their reductions, in Section
\S\ref{sec:overdensity} we present the comparison between the red
galaxy counts in this field and in deep field data, in Section
\S\ref{sec:RS} we present the full color-magnitude diagram and our
fits to the ``red sequence.''  We use a $(\Omega_{\Lambda},\Omega_{M})
= (0.73,0.27)$, $H_{0} = 71$ ${\rm km}$ ${\rm s^{-1}}$ ${\rm
  Mpc^{-1}}$ cosmology throughout.  At $z=2.16$ one arcsecond is
equivalent to 8.4 kpc.  All magnitudes are referenced to the AB system
\citep*{AB} unless otherwise noted.

\section{Observations, Data Reductions and Photometry\label{sec:obs}}

\begin{figure}
\plotone{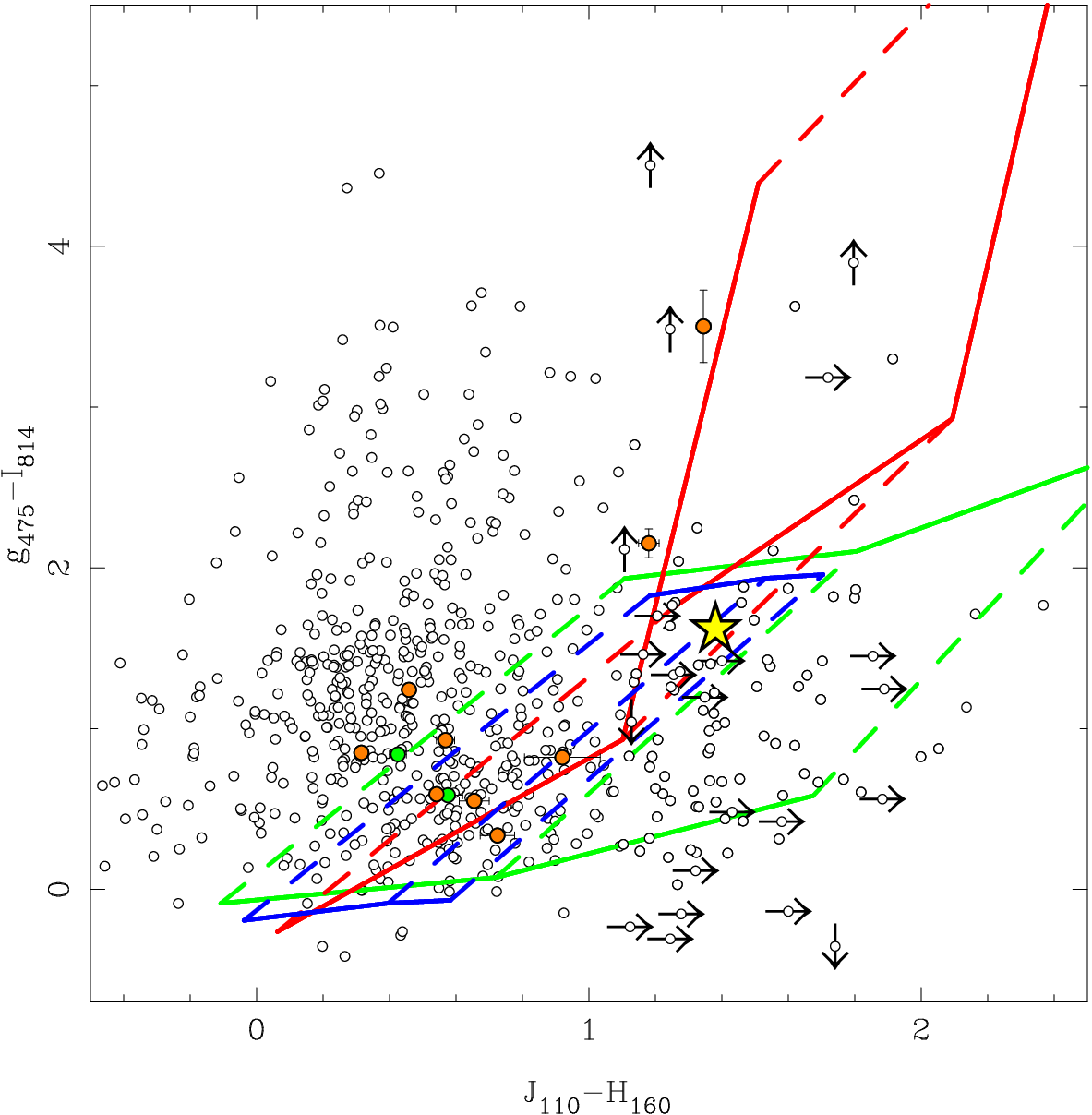}
\caption{$g_{435} - I_{814}$ vs. $J_{110}-H_{160}$ color-color
  diagram using the ACS and NICMOS data.  Arrows represent limits
  where the galaxy is only detected in a single band for that color.
  Filled circles indicate spectroscopically-confirmed Lyman-$\alpha$
  (green) and H-$\alpha$ (orange) emitting protocluster members. The
  yellow star is the radio galaxy.  The blue, green and red grids
  indicate the regions occupied by galaxies with an
  exponentially-decaying star-formation rate $\tau = 0.15$ Gyr (red),
  $\tau = 0.4$ Gyr (green) and a $\tau = 1000$ Gyr (blue) at $z=2.16$
  for three ages (0.1, 1 and 3 Gyr) and two different extinctions
  ($E(B-V) = 0.0, 1.0$).\label{fig:colorcolor}}
\end{figure}

The NICMOS instrument on-board {\it HST} is capable of deep
near-infrared imaging more quickly than from the ground but with a
relatively small field-of-view ($51\arcsec \times 51\arcsec$).  In the
case of MRC~1138-262, we know that galaxies are overdense on the scale
of a few arcminutes \citep*{Kurketal04, Croftetal05} and are thus
well-suited for observations with NICMOS camera 3 on {\it HST}.  We
used 30 orbits of {\it HST} time to image 10 of the 24 confirmed
members and $\sim 70$ of the candidate (narrow-band excess sources and
EROs) protocluster members in both the $J_{110}$ and $H_{160}$
filters.  We used seven pointings of NICMOS camera 3 in both filters
and one additional pointing in $H_{160}$ alone.  This single
`outrigger' $H_{160}$ pointing was included to obtain rest-frame
optical morphological information for a small concentration of
candidate members.  These observations reach an AB limiting magnitude
($m_{10\sigma}$; 10$\sigma$, $0\farcs5$ diameter circular aperture) of
$m_{10\sigma}=24.9$~mag in $J_{110}$ and $m_{10\sigma}=25.1$~mag in
$H_{160}$.  The same field was imaged in the $g_{475}$
($m_{10\sigma}=27.5$~mag) and $I_{814}$ ($m_{10\sigma}=26.8$~mag)
filters using the Wide-Field Channel of the Advanced Camera for
Surveys on {\it HST} as part of a Guaranteed Time program (\# 10327;
Miley et al. 2006\nocite{Mileyetal06}).  These optical data are useful
for their higher angular resolution and their coverage of the
rest-frame far-UV, thus extending the observed SEDs of candidate
protocluster members to shorter wavelengths where young stars and
on-going star-formation dominate the emitted spectrum.  In particular,
the $g_{475}$ and $I_{814}$ data allow us to partially differentiate
obscured star-formation from evolved stellar populations in the
candidate RS galaxies.

The NICMOS images were reduced using the on-the-fly reductions from
the {\it HST} archive, the IRAF task PEDSKY and the dither/drizzle
package to combine the images in a mosaic.  The dither offsets were
calculated using image cross-correlation and were refined with one
further iteration of cross-correlation.  Alignment of the pointings
relative to each other was accomplished using a rebinned version of
the ACS $I_{814}$ image as a reference.  The final mosaic has a pixel
scale of $0\farcs1$.  Galaxies were selected using the $H_{160}$-band
image for detection within SExtractor \citep*{SExtractor}.  We used a
$2.2\sigma$ detection threshold with a minimum connected area of 10
pixels.  We also corrected the NICMOS data for the count-rate
dependent non-linearity \citep*{CPSNONLINEAR}.  Total galaxy
magnitudes were estimated by using the MAG\_AUTO values from
SExtractor.

The $J_{110} - H_{160}$ colors were determined by running SExtractor
\citep*{SExtractor} in two-image mode using the $H_{160}$ image for
object detection and isophotal apertures.  The $J_{110}$ image was
PSF-matched to the $H_{160}$ band.  The resulting colors and
magnitudes are shown in Figure~\ref{fig:CMD}.  For galaxies which are
not detected at $2\sigma$ significance in the $J_{110}$-band (those to
the right of the thick dashed line, representing $J_{110, tot} >
26.7$, in Fig.~\ref{fig:CMD}) we consider the color to be a lower
limit.

We also measured similarly PSF-matched, isophotal colors using the two
ACS bands and have used them to construct a $g_{475}-I_{814}$ versus
$J_{110}-H_{160}$ color-color diagram (Figure~\ref{fig:colorcolor}).
We compared these colors to those of model SEDs for different ages,
star-formation histories and dust extinctions.  Using the 2007 Charlot
\& Bruzual\nocite{BC} population synthesis models we have constructed
spectral energy distributions for galaxies with an
exponentially-decaying star-formation rate with time constants of
$\tau = 0.15, 0.4, 1000.0$ Gyr (the red, green and blue grids in
Fig.~\ref{fig:colorcolor} respectively).  Each model's colors are
calculated for ages of 0.1, 1 and 3 Gyr and for $E(B-V) = 0.0$ and
$1.0$.  Aging of the population moves primarily the $J_{110}-H_{160}$
color to the red while the dust extinction significantly reddens the
$g_{475}-I_{814}$ color.  From this analysis it appears that the
$\tau=0.4$ Gyr model represents well the colors of a majority of the
red $J_{110}-H_{160}$ galaxies.

To extend the wavelength coverage for the protocluster galaxies we
also incorporated ground-based $U_{n}$-band data from LRIS-B on the
Keck telescope, $K_{s}$-band imaging from VLT/ISAAC and three band
IRAC imaging (the 3.6, 4.5 and 5.8 $\mu$m bands) from the {\it Spitzer
  Space Telescope}.  The Keck $U$-band data (PI W. van Breugel) were
obtained in late January and early February of 2003.  The ISAAC data
(PI G. Miley) were taken in Period 73 in service mode.  The {\it
  Spitzer} data are from the IRAC Guaranteed Time program (PI
G. Fazio, Program \#17).  We have smoothed the imaging data for all
bands, apart from the IRAC data, to match the resolution of the
$U_{n}$-band image (approximated by a FWHM~$\sim 1\arcsec$ Gaussian).
We then used SExtractor to measure galaxy magnitudes within a
$0\farcs5$ radius circular aperture for each smoothed image.  To
incorporate the IRAC data, which has much poorer angular resolution,
we derived aperture magnitudes which were then corrected to match the
smoothed data.  These aperture corrections were derived using the
photometric curves-of-growth for 20 stars in the field.  The resulting
catalog was used to generate photometric redshift estimates as
described below in Section \S~\ref{sec:photz}.

\begin{figure}
\plotone{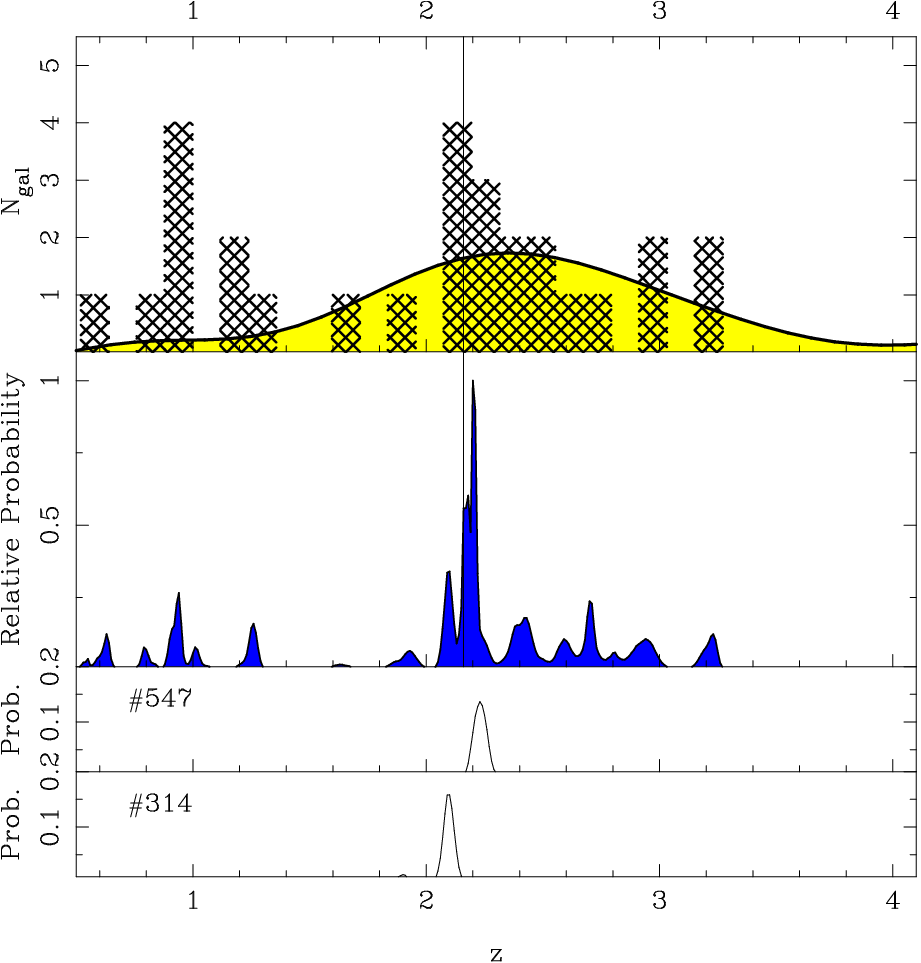}
\caption{upper panel: Distribution of high-confidence ($>95\%$)
  photometric redshifts and their selection function assuming a
  uniform $N(z)$ for our model template galaxies (yellow curve) for
  $1.1 \leq (J_{110}-H_{160}) \leq 2.1$ galaxies in the MRC 1138-262
  NICMOS field.  The peak between $z=2.1$ and $z=2.4$ is statistically
  highly significant.  middle panel: Sum of redshift probability
  distributions for all the galaxies in the upper panel.  38.5\% of
  the total probability is contained in the redshift interval from 2
  to 2.3. lower 2 panels: two examples of the probability distribution
  function for individual galaxies\label{fig:photzRS}}
\end{figure}

\section{Photometric Redshifts\label{sec:photz}}

We have used the ACS ($g_{475}$, $I_{814}$), NICMOS ($J_{110}$,
$H_{160}$), ground-based $U_{n}$-band from Keck/LRIS-B, $K_{S}$-band
imaging from VLT/ISAAC and {\it Spitzer}/IRAC imaging to estimate
photometric redshifts for our $H_{160}$-band selected sample.  We
input a catalog of aperture galaxy magnitudes, based on the matched,
smoothed images described above, into the Bayesian photometric
redshift code (BPZ) of Ben{\'{\i}}tez (2000)\nocite{BPZ} using a
uniform prior.  We felt that the default prior, based on optical
galaxy selection and spectroscopy in the HDF-N, would not necessarily
represent the redshift distribution for our near-infrared selected
galaxies.  We generated our own extensive set of template spectral
energy distributions using the models of Charlot \& Bruzual
(2007)\nocite{CB07}.  All these SEDs are $\tau$ models with values for
$\tau = [0.15, 0.4, 1.0, 2.0, 1000.0]$ Gyr and ages $=[0.05, 0.1, 0.5,
1.0, 2.0, 3.0]$ Gyr.  We also included models with internal dust
extinction ranging from $E(B-V) = [0.0, 0.1, 0.3, 0.5, 0.75, 1.0]$ mag
and metallicity of $(Z/Z_{\odot}) = [0.3, 1.0, 2.5]$.
We focused particular attention on the $J_{110}-H_{160}$ selected
surface-overdensity.  In the upper panel of Figure~\ref{fig:photzRS}
we present the high confidence ($> 95\%$) photo-$z$ distribution for
the NIR-color selected ($1.1 \leq (J_{110} - H_{160}) \leq 2.1$)
subsample.  We ran extensive simulations by redshifting our template
set, adding appropriate photometric errors and using BPZ to recover
the redshifts.  The yellow curve represents the redshift selection
function for this color cut, template set and filters assuming that
these model galaxies follow a uniform $N(z)$ over this redshift
interval.  The simulation results were free of significant systematic
errors and the random errors are estimated to be $\delta z/z \sim
0.1$.  Based on these SED fits, the approximate luminosity-weighted
ages of the red galaxies lie between 1 and 2.5 Gyrs and their stellar
masses are of order a few $\times 10^{10} M_{\odot}$.  These stellar
masses are reasonable as are the absolute magnitudes (see
Figure~\ref{fig:CMRfits}).  More detailed SED modeling is deferred to
a future paper.

There is a clear excess of galaxies between $z=2.0$ and $z=2.5$.  For
each galaxy fit by BPZ we have generated the full redshift probability
distribution.  In the lower panel of Fig.~\ref{fig:photzRS} we show
the $H_{160}$-band weighted-average of these probability
distributions.  There is a clear peak (containing 38.5\% of the total
probability compared to only 17\% of the total selection function in
the same redshift interval) between $z=2.0$ and $z=2.3$, consistent
with the significant peak in the redshift histogram itself.

\section{NICMOS Galaxy Morphologies\label{sec:morf}}

\begin{figure}[b]
\plotone{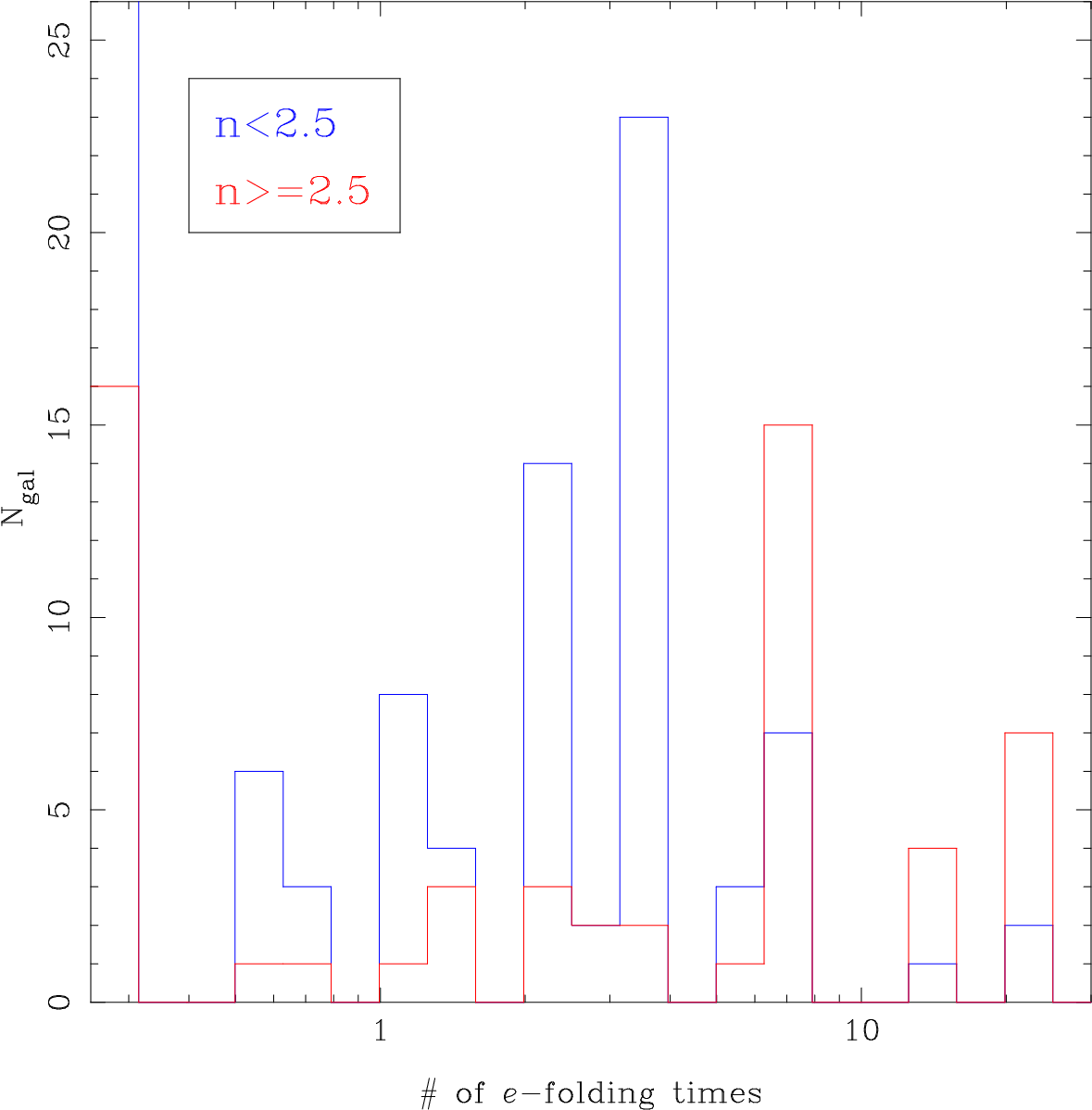}
\caption{Distribution of inferred stellar ages (in terms of $\tau$)
  for both the concentrated ($n\geq2.5$, red line) and diffuse
  ($n<2.5$, blue line) galaxies which are well-resolved in the NICMOS
  data.  Constant star-formation models, for which the $e$-folding
  time is infinite, are placed at the left-hand edge of the plot.  The
  blue and red distributions are quite different.  Of particular note
  is that the most evolved galaxies generally have high $n$ while the
  low $n$ galaxies dominate the star-forming
  population.\label{fig:tAn}}
\end{figure}

NICMOS camera 3 provides good angular resolution over its entire
field-of-view.  The FWHM of the PSF in our final mosaic is $\approx
0\farcs27$.  To exploit this resolution we have used the GALFIT code
\citep*{GALFIT} to fit analytic S{\'e}rsic surface-brightness profiles
\citep*{Sersic} to all the $H_{160} \leq 24.5$ sources in our
$H_{160}$-band mosaic.  A model point-spread function was created for
each of these galaxies individually by generating a TinyTim simulated
PSF \citep*{TINYTIM} at the galaxies' positions in each exposure and
then drizzling these PSFs together in exactly the same fashion as for
the data themselves (see Zirm et al. 2007\nocite{Zirmetal07}).  We
restricted the S{\'e}rsic index, $n$, to be between 1 and 5.  We will
present a full analysis of the morphologies of these galaxies in a
future paper.  For the current work, we use these derived sizes and
profile shapes to assist us in selecting the morphological
``early-type'' members of the red galaxy population.  

\begin{figure*}[t]
\plotone{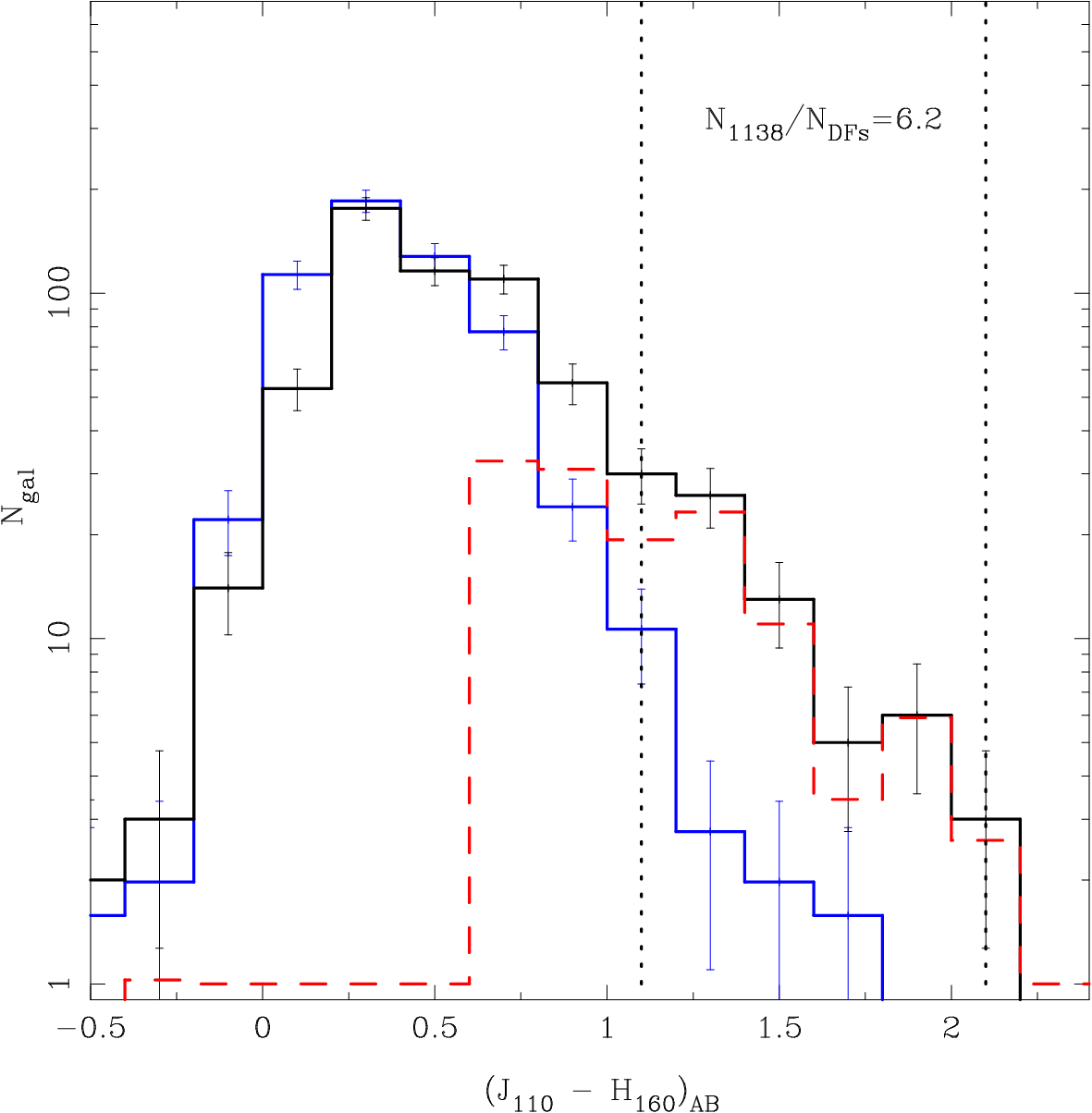}
\caption{Histogram of the color distributions for the 1138 and deep
  fields (blue).  The deep field data has been normalized by total
  area to the 1138 data.  Note the clear excess of red galaxies in the
  1138 field. At $1.1 \leq (J_{110} - H_{160}) \leq 2.1$ (horizontal
  dotted lines) for galaxies brighter than the 2$\sigma$
  $J_{110}$-band limit (dashed line) there is an overdensity of a
  factor of 6.2 in the 1138 field.  \label{fig:hist}}
\end{figure*}

In Figure~\ref{fig:tAn} we show the distribution of galaxy ages
derived via these SED fits as parametrized by the $\tau$ value for
the best-fitting model for those galaxies with high and low S{\'e}rsic
index ($n \geq 2.5$, red line, and $n < 2.5$, blue line).  It is clear
that while there is substantial overlap between these distributions
they are not identical and that they differ in the sense that one
might expect, namely, that the concentrated galaxies appear to be
comprised of older stellar populations.  This trend gives us some
confidence in trying to select the ``early-type'' galaxies using these
data which is important for our discussion of the color-magnitude
relation in Section~\ref{sec:RS}.
	
\section{Surface Overdensity of Red Galaxies\label{sec:overdensity}}

\begin{figure*}[t]
\plotone{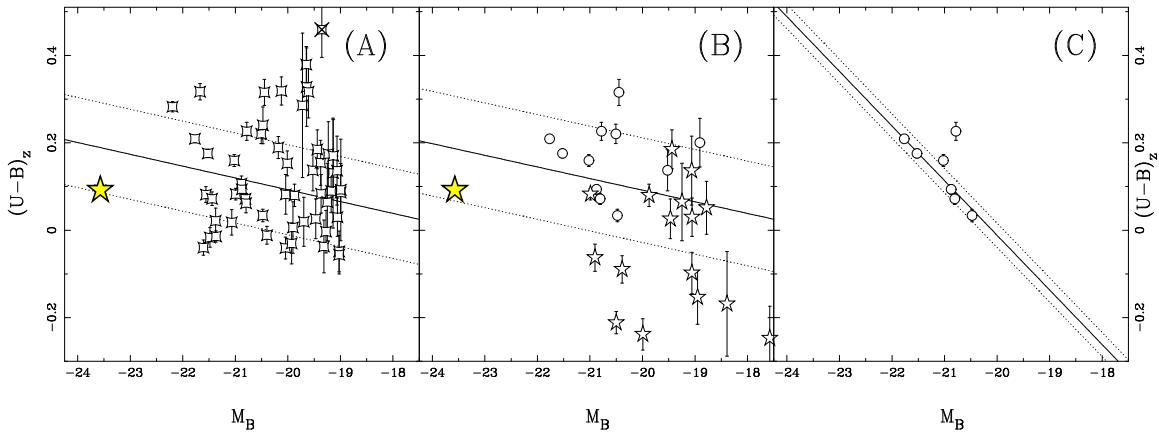}
\caption{Linear fits to the rest-frame $U-B$ (Vega) color-magnitude
  diagrams for three different sub-samples of the $H_{160}$-band
  selected NICMOS sample.  Panel (A) shows the fit (solid line) and
  intrinsic scatter ($1\sigma$; dotted lines) for a sample selected to
  have $1.1 \leq (J_{110} - H_{160}) \leq 2.1$.  The crossed-out point
  are those which are rejected as outliers in more than half of the
  realizations (see \S\ref{sec:RS}).  Both the observed and intrinsic
  scatter are smaller than the initial color cut.  Panel (B) shows the
  fit and intrinsic scatter for a photometric redshift selected sample
  with $2.0 < z_{\rm phot} < 2.5$.  The stars indicate galaxies whose
  preferred photometric template has an age $< 4\tau$, while circles
  represent galaxies with older than this limit.  Panel (C) shows the
  fit and scatter for those galaxies which meet the same redshift cut
  but also are well-resolved with a high S{\'e}rsic index ($n > 2.5$)
  and best-fit by an age $\geq 4\tau$ template.\label{fig:CMRfits}}
\end{figure*}

To compare this protocluster field to more generic `blank' field data
we have compiled catalogs for the public NICMOS data in both the
Hubble Deep Field North (HDF-N) and the Ultra Deep Field (UDF).
Figure~\ref{fig:CMD} shows the $J_{110} - H_{160}$ color-magnitude
diagram (open black circles) and the color distributions for both the
MRC~1138-262 and the combined HDF-N and UDF galaxy catalogs (blue
circles).  The deep field data were also $H_{160}$-band selected.  The
area of the two deep fields is roughly 2.5 times the area of our
protocluster observations.  We have applied no correction to the deep
field number counts to account for clustering in those fields.  The
color histogram in Figure~\ref{fig:hist} shows the area-normalized
galaxy counts from the two deep fields (blue line) and from the 1138
field to the same ($2\sigma$) limiting magnitude of $J_{110} =
26.7$~mag (AB).  The red dashed line shows the difference between the
two color distributions.  It is clear that the radio galaxy field is
overdense in red galaxies by a large factor.  For sources with colors
between $1.1 \leq (J_{110} - H_{160}) \leq 2.1$, the
horizontal(vertical) dotted lines in
Fig.\ref{fig:CMD}(\ref{fig:hist}), and brighter than our
$J_{110}$-band $2\sigma$ limit ($26.7$), we calculate an
area-normalized overdensity of $6.2$ when compared to the deep fields
data.  We note that the exact value of the measured overdensity is
rather sensitive to systematic color offsets between the protocluster
and deep field data.  A redward shift of 0.05 for the deep field
galaxies would lower the measured overdensity to $5.0$.  However, we
are confident that these systematic offsets remain small ($< 0.05$
mag) since we have used the same instrument, filters, selection
technique and photometric code with very similar input parameters for
both the deep field and 1138 datasets.  Looking back at
Figure~\ref{fig:colorcolor} we can see that many of the
spectroscopically-confirmed line emitters (filled blue circles) and
red galaxies in the overdensity are well-represented by the $\tau =
0.4$ Gyr model (green lines) at different ages and extinctions.

This current work is not the first to observe red galaxies in this
field.  Kurk et al. (2004)\nocite{Kurketal04} identified a small
($\sim 1.5\times$) surface overdensity of extremely red objects (EROs;
$I-K > 4.3$ Vega magnitudes) using ground-based $I$ and $K$ band data.
Many of these EROs are also identified as red in the NICMOS
$J_{110}-H_{160}$ color.  More recently Kodama et
al. (2007)\nocite{Kodamaetal07} observed this field using the
wide-field NIR imager, MOIRCS, on the Subaru telescope.  These authors
found several bright (presumably massive) red galaxies over a wider
field-of-view but to shallower depths than the NICMOS data presented
here.  24 of their color-selected protocluster candidates are within
our NICMOS mosaic.  23 of the 24 are identified in our data as being
red in $J_{110} - H_{160}$.  Furthermore, 18 of the 94 galaxies which
satisfy our color criteria (and have $J_{110}<26.67$) are also
identified by Kodama et al. as protocluster candidates.  The much
larger number of red galaxies in the NICMOS data is primarily due to
fainter galaxies detected at high significance in our deeper data.

\section{An Emergent Red-Sequence?\label{sec:RS}}

\begin{figure*}[t]
\epsscale{0.8}
\plotone{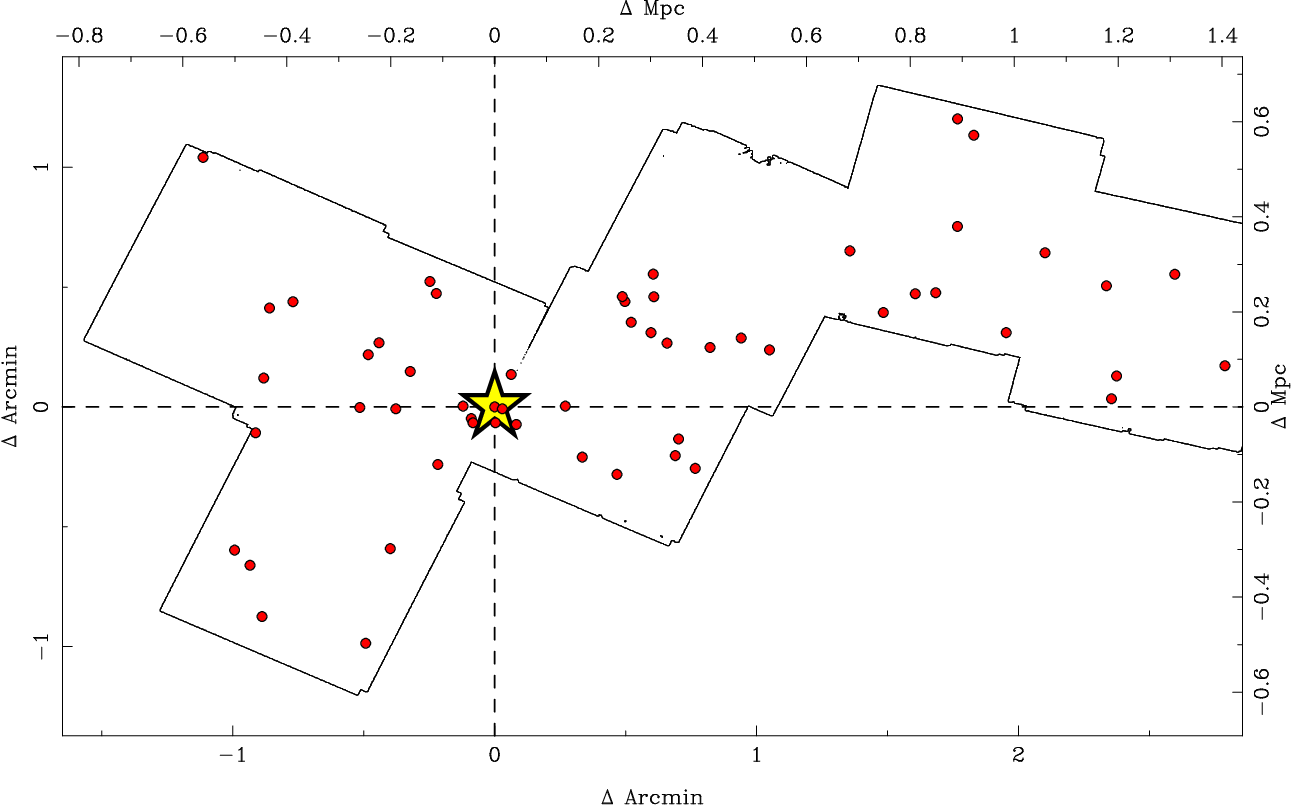}
\caption{Spatial distribution of galaxies (red circles) relative to
  the radio galaxy MRC~1138-262 (yellow star) for the color-selected
  sample defined in the text (A) and shown in the first panel of
  Fig.~\ref{fig:CMRfits}.  The irregular black outline encloses the
  coverage of the 7 NICMOS camera 3 pointing with both $J_{110}$ and
  $H_{160}$ imaging.\label{fig:spatial}}
\end{figure*}

To study the colors and magnitudes of these galaxies in more detail
and to possibly identify a red-sequence in the 1138 field we have
split the galaxies into three sub-samples defined by $J_{110}-H_{160}$
color, photometric redshift and morphology (S{\'e}rsic index).  The
first sample (Sample A) comprises all 56 galaxies with $1.1 \leq
(J_{110}-H_{160}) \leq 2.1$ and $H_{160} < 24.5$ and includes the
radio galaxy itself.  Sample B is made up of all 28 galaxies with a
robust photometric redshift between 2.0 and 2.5 and $J_{110}-H_{160} >
0.75$ and $H_{160} < 26.0$.  This liberal color cut is included to
select galaxies which comprise the large observed surface-overdensity.
Finally, sample C contains seven galaxies with the same photometric
redshift cut but which also have well-resolved $H_{160}$-band
surface-brightness profiles with S{\'e}rsic index $n > 2.5$.  All of
these galaxies' SEDs are also best-fit by models with relatively
little on-going star-formation.  We use a limit of (age $\geq 4 \times
\tau$, cf. Grazian et al 2007 \nocite{Grazianetal07}).  Therefore,
sample C mimics the color, morphological and photometric redshift
selection of early-type galaxies in clusters at $z \simlt 1$.  The
photometry, photo-$z$s and sizes of the sample C galaxies are listed
in Table~\ref{tab:RS}, their rest-frame color-magnitude distributions
are shown in Figure~\ref{fig:CMRfits} and the two-dimensional spatial
distribution of the Sample A galaxies is plotted in
Figure~\ref{fig:spatial}.  We note that because the measured
overdensity is a factor of 6, we statistically expect one of every
seven sample A galaxies to be a field galaxy.  However, this should
not effect our results significantly.

For these three sample selections we have fit a line and measured the
intrinsic scatter about that best-fit line (see
Fig.~\ref{fig:CMRfits}).  For comparison to lower redshift galaxy
clusters we have transformed our observed $J_{110}-H_{160}$ color and
$H_{160}$ magnitudes into rest-frame $U-B$ and $B$ (Vega),
respectively, using the following expressions:
\begin{equation}
(U-B)_{\rm rest} = 0.539 \times (J_{110}-H_{160})_{\rm obs} - 0.653
\end{equation}

\begin{equation}
  M_{B, {\rm rest}} = H_{160,{\rm obs}} - 0.170 \times (J_{110}-H_{160})_{\rm obs} - 43.625
\end{equation}

The small color corrections used in these relations were derived using
a family of $\tau$-models with a range of ages (0.1-12 Gyr), $\tau$
(0.1-5 Gyr) and three metallicities ($0.4, 1$ and $2.5 Z_{\odot}$).

To fit the ``CMR'' we used a bootstrap re-sampling technique to
estimate the error on the fitted slope.  Then, by assuming that all
the red galaxies lie on this fit line, we ran Monte Carlo realizations
of the contribution of the photometric errors to the observed color
scatter about the fit line, i.e., by fixing a color-magnitude relation
we calculate the measurement scatter with zero intrinsic scatter.  We
then calculate the intrinsic scatter by subtracting (in quadrature)
the estimated measurement scatter from the observed scatter.  We show
these fits (solid line) and the intrinsic scatters (dotted lines) for
the three samples (A, B and C) in Figure~\ref{fig:CMRfits}.  The fits
to both Sample's A and B have nearly identical rest-frame $U-B$
slopes, $0.027$ and $0.026$ respectively, and intrinsic scatters
($0.10$ and $0.12$).  While these slopes are comparable to those found
for the well-populated lower redshift cluster CMRs, the intrinsic
scatters are considerably higher.  However, the scatter measured for
the eight galaxy Sample C is comparable to that of the lower redshift
samples but with a much steeper slope ($0.130$).  When these scatters
are compared to model predictions based on lower redshift clusters
(specifically RDCS 1252.9-2927 at $z=1.24$; Gobat et al. 2008) we find
that the 1138 protocluster has lower than predicted scatter.  This may
suggest that the 1138 protocluster is in a more advanced evolutionary
state than RDCS 1252 was at $z=2.2$.

We have calculated three representative color-magnitude relations for
comparison to the colors and magnitudes of the red galaxies (three
dot-dash lines in Fig.~\ref{fig:CMD}).  We have taken two lower
redshift clusters, Coma at $z=0.023$ and RDCS~1252.9-2927 at $z=1.24$,
and transformed them to the observed filters and $z=2.16$ under the
assumption that the colors do not evolve.  In this no evolution case
(the two dot-dash lines in Fig.~\ref{fig:CMD}) the CMRs appear at the
red edge of the observed overdensity.  There is almost exactly 2 Gyr
of cosmic time between $z=2.16$ and $z=1.24$ in our adopted cosmology.
From \citet{Blakesleeetal03} we know that the median redshift of last
significant star-formation for the RDCS~1252 galaxies is between
$z=2.7-3.6$.  Therefore, if we observe those galaxies at $z=2.16$ they
will be significantly younger and hence bluer.  In fact, this
passively de-evolved line (bluest dot-dash line in Fig.~\ref{fig:CMD},
labeled `$z_{\rm form} \sim 3$') does fall blueward of the red galaxy
overdensity.  We discuss the implications for these comparisons in
Section~\ref{sec:conclusions}.

We have also translated the Kodama et al. ground-based $J-K$ colors to
our NICMOS filters assuming all the red galaxies lie at $z=2.16$.
These bright galaxies also fall along the passively de-evolved line
with the radio galaxy.  We have used our suite of SED models to
estimate the color transformation from their ground-based $J-K_{S}$ to
our NICMOS $J_{110}-H_{160}$ color.  Roughly, the Kodama et al. bright
red galaxies fall where the RDCS~1252 passive line crosses our color
cut at $J_{110}-H_{160} = 1.1$.  This result hints at a possible
bi-modality in the red galaxy population of this protocluster.
Namely, that there are faint red galaxies that are inconsistent with
passively-evolving cluster members either due to large amounts of
dust, or due to higher redshifts of formation but that the more
luminous protocluster members may have already finished forming and
seem consistent with passive evolution to the present-day.

\section{Discussion\label{sec:conclusions}}

We have identified a (6.2$\times$) surface-overdensity and a
corresponding photometric redshift `spike' of red $J_{110}-H_{160}$
galaxies which are likely associated with a known protocluster at
$z=2.16$.  The optical-NIR spectral energy distributions of these
sources suggest that they comprise both evolved galaxies as well as
dust-obscured star-forming galaxies.  Based on our SED fits from the
photometric redshift determinations, the approximate
luminosity-weighted ages of these sources lie between 1 and 2.5 Gyrs
and their stellar masses are of order a few $\times 10^{10}
M_{\odot}$.  Detailed modeling of the SEDs for the protocluster
population, along with their morphologies, is reserved for a future
paper.

Comparison with the CMRs of lower redshift clusters shows that the red
galaxy overdensity primarily lies blueward of the no-evolution
predictions.  That the red galaxies in 1138 are also redder than the
$z_{\rm form} \sim 3$ case suggests both that there are galaxies with
significant dust content, an assertion supported by the SED fits, and
also that they were perhaps formed at higher redshift than the
RDCS1252 galaxies.  Of course, without a classical, low-scatter
red-sequence to use as a baseline there remains considerable
uncertainty in the age of the population as a whole.  The results of
Steidel et al. (2005)\nocite{Steideletal05} suggest that protocluster
galaxies are older than their ``field'' counterparts at $z \sim 2.3$
and that these ages and stellar masses were broadly consistent with
evolution to lower redshift cluster galaxies.  However, their
protocluster members were all UV-selected and star-forming.  With
future spectroscopy of our red galaxy sample it will be possible to
see if these differences persist when looking at a more varied galaxy
sample.

For three samples of galaxies drawn from the full $H_{160}$-band
selected dataset we have fit a color-magnitude relation and estimated
the intrinsic scatter.  The CMR at $z=2.16$ is not as well-defined as
at $z \sim 1$ or 0.  For sample C, made up of 8 galaxies, the color,
best-fit spectral template, morphology and photo-$z$ all point towards
them being (proto-)elliptical galaxies within the protocluster.  For
this small sample, the estimated intrinsic scatter is rather low and
may suggest that these galaxies represent the forming red-sequence in
this protocluster.  The slope of this relation is extremely steep
compared to lower redshift clusters.  The slope of the CMR is
generally assumed to be a manifestation of the mass-metallicity
relation and would therefore flatten at higher redshift.  The major
caveat regarding the steep slope of Sample C is that none of these
galaxies are spectroscopically confirmed protocluster members.
Therefore, this ``relation'' may just be a random, although somewhat
unlikely, coincidence rather than a nascent CMR.  However, further
deep NIR imaging coverage of this field is required to identify
additional members of this proto-elliptical galaxy class.

\acknowledgments

Support for program \# 10404 was provided by NASA through a grant
(GO-10404.01-A) from the Space Telescope Science Institute, which is
operated by the Association of Universities for Research in Astronomy,
Inc., under NASA contract NAS 5-26555.  The work of SAS was performed
in part under the auspices of the U.S. Department of Energy, National
Nuclear Security Administration by the University of California,
Lawrence Livermore National Laboratory under contract
No. W-7405-Eng-48.  JK is financially supported by the DFG, grant SFB
439.  WvB acknowledges support for radio galaxy studies at UC Merced,
including the work reported here, with the Hubble Space Telescope and
the Spitzer Space Telescope via NASA grants HST \# 10127, SST \#
1264353, SST \# 1265551, SST \# 1279182.


\tabletypesize{\tiny}
\begin{deluxetable*}{lccccccccccc}[b]
\tablecolumns{12}
\tablewidth{0in}
\tablecaption{Properties of Red Galaxies in the MRC 1138-262 Field: $ H_{160} < 24.5 \wedge 1.1 \leq (J_{110} - H_{160}) \leq 2.1$\label{tab:RS}}
\tablehead{
\colhead{ID} & \colhead{$z_{\rm phot}$\tablenotemark{a}} & \colhead{Odds\tablenotemark{b}} & \colhead{SED Type\tablenotemark{c}} & \colhead{S{\'e}rsic} & \colhead{$r_e$} & \colhead{$r_e$} & \colhead{$H_{160}$\tablenotemark{d}} & \colhead{$U_{n} - g_{475}$}  & \colhead{$g_{475} - I_{814}$}  & \colhead{$I_{814} - J_{110}$} & \colhead{$J_{110} - H_{160}$}\\
\colhead{} & \colhead{} & \colhead{(BPZ)} & \colhead{} & \colhead{Index} & \colhead{($\arcsec$)} & \colhead{(kpc)} & \colhead{(AB)} & \colhead{(AB)} & \colhead{(AB)} & \colhead{(AB)} & \colhead{(AB)}\\
\colhead{} & \colhead{} & \colhead{} & \colhead{} & \colhead{($n$)} & \colhead{} & \colhead{} & \colhead{} & \colhead{} & \colhead{} & \colhead{} & \colhead{}}
\startdata
606 & 2.04 & 0.98 & 1.00 & 3.4 $\pm$ 0.4 & 0.21 $\pm$ 0.01 & 1.79 $\pm$ 0.10 & 23.09 $\pm$ 0.03 & 1.10 & 0.79 & 1.18 & 1.06 \\ 
507 & 2.09 & 1.00 & 1.00 & 5.0 $\pm$ 0.4 & 0.24 $\pm$ 0.01 & 1.99 $\pm$ 0.08 & 22.50 $\pm$ 0.01 & 28.88 & -27.16 & 2.22 & 1.52 \\ 
314 & 2.10 & 1.00 & 1.00 & 5.0 $\pm$ 0.7 & 0.15 $\pm$ 0.01 & 1.28 $\pm$ 0.05 & 23.03 $\pm$ 0.01 & 1.92 & 1.38 & 1.89 & 1.29 \\ 
586 & 2.10 & 1.00 & 1.00 & 5.0 $\pm$ 0.3 & 0.40 $\pm$ 0.03 & 3.34 $\pm$ 0.23 & 21.77 $\pm$ 0.01 & 1.80 & 1.89 & 2.22 & 1.61 \\ 
547 & 2.23 & 1.00 & 1.00 & 5.0 $\pm$ 0.6 & 0.12 $\pm$ 0.00 & 1.02 $\pm$ 0.04 & 22.70 $\pm$ 0.01 & 1.30 & 1.13 & 1.67 & 1.36 \\ 
493 & 2.25 & 1.00 & 1.00 & 3.5 $\pm$ 0.3 & 0.56 $\pm$ 0.04 & 4.72 $\pm$ 0.36 & 22.75 $\pm$ 0.02 & 0.63 & 1.28 & 2.19 & 1.64 \\ 
312 & 2.42 & 1.00 & 1.00 & 5.0 $\pm$ 0.3 & 0.38 $\pm$ 0.02 & 3.20 $\pm$ 0.19 & 22.00 $\pm$ 0.01 & -28.02 & 1.45 & 1.94 & 1.55 \\ 
127 & 2.44 & 1.00 & 1.00 & 2.7 $\pm$ 0.6 & 0.12 $\pm$ 0.02 & 1.01 $\pm$ 0.18 & 22.64 $\pm$ 0.01 & 0.34 & 1.04 & 1.37 & 1.40 \\ 
\enddata
\tablenotetext{a}{The errors on these photometric redshifts are
  approximately 0.1-0.2} 
\tablenotetext{b}{The ODDS parameter approaches unity when the probability distribution function has a single narrow peak}
\tablenotetext{c}{Best-fit broad-band template type: 0: dusty and
  star-forming 1: quiescent} 
\tablenotetext{d}{Total magnitudes based
  on the SExtractor MAG\_AUTO parameter}
\end{deluxetable*}

\end{document}